\documentclass[reprint,twocolumn,showkeys,superscriptaddress]{revtex4}

\usepackage{graphicx}
\usepackage{dcolumn}
\usepackage{bm}
\usepackage{color}
\usepackage{amsmath,amsthm}
\usepackage{amssymb,bm}
\usepackage{mathrsfs}
\usepackage{comment}
\usepackage{ulem}

\newcommand{\ket}[1]{|#1\rangle}
\newcommand{\bra}[1]{\langle #1|}

\begin{document}

\title{Variational Polaron Theory for Ground States of Strongly Coupled Light--Matter and Electron--Phonon Systems}
\author{Nguyen Thanh Phuc}
\email{nthanhphuc@moleng.kyoto-u.ac.jp}
\affiliation{Department of Chemical Science and Engineering, Graduate School of Engineering, Kyoto University, Kyoto 615-8510, Japan}

\begin{abstract}
	Strong light--matter and electron--phonon coupling generate ground states dressed by virtual bosonic excitations, making bare-state truncations and perturbative treatments unreliable in the ultrastrong-coupling regime. 
	We introduce a nonperturbative variational ground-state framework based on a state-dependent polaron transformation, combined with a product-state ansatz and a second-order perturbative correction for residual matter--boson entanglement. 
	We show that the optimized transformed frame becomes asymptotically decoupled at infinite coupling, because the leading linear coupling is canceled while off-diagonal matter transitions are suppressed by displaced-oscillator overlaps. 
	The approach is asymptotically correct in both weak- and strong-coupling limits and remains accurate in the intermediate regime, where fixed polaron transformations are least reliable. 
	Dicke-model benchmarks reproduce ground-state energies, fidelities, and the superradiant transition, with second-order energy errors below 0.2\%. 
	Holstein-model benchmarks yield errors below 0.5\% and clarify how translational symmetry affects wave-function quality. 
	This dressed-basis framework enables nonperturbative modeling of strongly coupled light--matter and electron--phonon systems.
\end{abstract}


\maketitle

\section*{TOC Graphic}
\begin{center}
	\includegraphics[width=3.25in,height=1.75in]{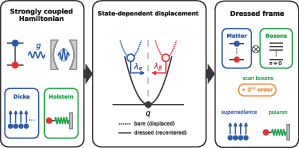}
\end{center}

Strong coupling between matter degrees of freedom and bosonic modes is a central theme in molecular science, condensed-matter physics, and cavity quantum electrodynamics. 
In optical cavities, electronic and vibrational excitations can hybridize with confined electromagnetic modes to form molecular polaritons, enabling new ways to modify molecular spectra, energy flow, and chemical reactivity~\cite{Ebbesen2016,Ribeiro2018,Herrera2020,GarciaVidal2021}. 
Recent theoretical and experimental studies have shown, for example, that polariton formation can sharpen spectroscopic signatures and suppress environmental decoherence~\cite{Phuc2019PRResearch, Spano2015, Herrera2016, Takahashi2020}, modify electron-transfer rates through strong light-matter coupling~\cite{Semenov2019, Angulo2019, Phuc2020, Mandal2020, Chowdhury2022, Wei2022, Saller2023, Ying2026, Hidaka2026}, enhance charge or excitation-energy transfer through collective and Bose-stimulated mechanisms~\cite{Phuc2021JCP, Phuc2022, Koessler2025}, and enable optical control of spin selectivity and transport in photon-coupled or periodically driven molecular systems~\cite{Phuc2023JPCL, Phuc2023JCP, Liu2025, Phuc2018, Phuc2019PRB, Wang2024}. 
In molecular and organic materials, charge carriers can become strongly dressed by local vibrational modes through electron--phonon coupling, giving rise to polarons and polaronic transport~\cite{Holstein1959,LangFirsov1963,Mahan2000,Alexandrov2010}. 
These examples share a common physical feature: the relevant low-energy states are not simple products of bare matter and bosonic states, but dressed states containing virtual photonic or vibrational excitations. 
Such dressing provides the microscopic origin of many cavity- and phonon-induced modifications  of molecular and material properties, but it also makes the ground-state problem intrinsically nonperturbative.

The difficulty becomes especially pronounced in the ultrastrong- and deep-strong-coupling regimes, where the coupling strength is comparable to, or even exceeds, the characteristic bosonic frequency or matter excitation energy~\cite{Kockum2019,FornDiaz2019}. 
In these regimes, the rotating-wave approximation is no longer reliable, counter-rotating processes contribute directly to the ground state, and bare-state truncations of the bosonic Hilbert space can become inefficient or uncontrolled. 
Weak-coupling perturbation theory fails because the bosonic dressing is no longer small, whereas conventional strong-coupling polaron pictures may become quantitatively inaccurate in the intermediate regime. 
A useful ground-state theory should therefore interpolate between weak and strong coupling, remain accurate where neither limiting description is sufficient, and provide a compact physical representation of matter--boson dressing.

Several recent works have addressed related aspects of this problem. 
Exponential ansatzes and variational Lang--Firsov transformations have been developed for Holstein and ab initio polariton problems~\cite{Yang2024,Cui2024}, while molecular quantum-electrodynamical and cavity-QED molecular orbital theories have clarified electronic-structure modifications by quantized radiation fields~\cite{LiZhang2023,Riso2022}. 
Variational polaron transformations are also central to open quantum systems and nonperturbative cavity-QED descriptions~\cite{SilbeyHarris1984,DeLiberato2014,Ashida2021}. 
These transformed-frame approaches are complementary to scalable dynamical methods such as truncated-Wigner simulations of molecular polariton dynamics~\cite{Phuc2024, Phuc2025}.

Despite these important advances, a complementary ground-state strategy remains desirable. 
First, many transformed-frame mean-field or electronic-structure approaches use a transformed bosonic vacuum, coherent state, or similarly restricted bosonic reference as the zeroth-order state~\cite{Cui2024,LiZhang2023,Riso2022}. 
This choice is computationally attractive, but it does not explicitly scan the residual bosonic Hilbert space and can miss residual bosonic excitations beyond the transformed vacuum that survive after an incomplete dressing transformation. 
Second, although previous studies employ polaron, Lang--Firsov, or related exponential transformations, they do not generally make explicit or exploit the asymptotic decoupling of the transformed Hamiltonian in the infinite-coupling limit as the organizing principle of the variational ground-state theory.
Such an asymptotic decoupling property is useful because it provides a controlled strong-coupling reference: in the transformed frame, the residual matter--boson coupling vanishes as the coupling strength becomes infinite, and a product-state ansatz becomes asymptotically exact.
Conversely, the asymptotic-decoupling transformation developed for cavity QED is nonvariational in the present sense: its transformation is not optimized by minimizing the finite-coupling ground-state energy for each value of the coupling strength~\cite{Ashida2021}.
The present method combines these two ingredients: an asymptotically decoupled transformed frame and variationally optimized displacement parameters at finite coupling.

A further practical issue is that asymptotic decoupling is a limiting property, not a sharp transition at a universal finite coupling. 
Applications require quantitative diagnostics of how rapidly the residual transformed-frame coupling, energy error, and wave-function infidelity decrease with coupling strength. 
Such diagnostics are needed to decide when a compact dressed-basis description is reliable, when non-vacuum bosonic components must be retained, and when residual matter-boson correlations require perturbative correction.

Building on these ideas, we introduce a nonperturbative variational ground-state framework based on a state-dependent polaron transformation. 
We show generally that the optimized transformed frame becomes asymptotically decoupled in the infinite-coupling limit: the optimized displacement cancels the leading diagonal linear matter--boson coupling, while off-diagonal matter-sector transitions are suppressed by displaced-oscillator overlaps. 
This asymptotic decoupling is used not only as a formal strong-coupling statement, but also as a physical basis and quantitative diagnostic for when a compact dressed-basis description becomes reliable.
Although finite-coupling systems still retain residual matter-boson correlations, the strong-coupling limit suggests that a properly optimized dressed frame should make these correlations perturbative. 
It is therefore natural to start from a zeroth-order state that factorizes between the matter and bosonic sectors, and then incorporate the remaining correlations through a second-order perturbative correction.
At finite coupling, the displacement parameters are optimized variationally by minimizing the ground-state energy, allowing the dressed basis to interpolate between the bare weak-coupling description and the asymptotically decoupled strong-coupling description.

Importantly, the product-state reference is factorized only between the matter and bosonic sectors. 
The bosonic component is not restricted to the transformed vacuum, nor is it assumed to factorize over independent modes. 
Instead, it is expanded over the retained multimode bosonic Hilbert space and optimized together with the matter component and the displacement parameters. 
This structure also provides a practical advantage: because the matter and bosonic components are separated at the zeroth-order level, the matter sector can in principle be treated using problem-specific methods developed for correlated matter systems, while the bosonic dressing is handled in the optimized transformed frame. 
Residual matter-boson entanglement beyond this optimized product state is then incorporated by second-order perturbation theory.

We benchmark the method using two paradigmatic models that represent complementary types of strongly coupled systems. 
The Dicke model serves as a minimal model of collective light--matter coupling between many two-level emitters and a single cavity mode. 
Although it is not a complete molecular polariton Hamiltonian because it does not explicitly include nuclear or vibronic structure, it provides a stringent test for counter-rotating light--matter coupling, collective enhancement, and the superradiant quantum phase transition. 
The Holstein model, by contrast, describes itinerant electronic degrees of freedom coupled to local vibrational modes and is directly relevant to polaron formation in molecular and organic materials. 
By benchmarking against exact diagonalization, we quantify both the practical onset of the effectively decoupled regime and the residual energy and fidelity errors throughout the intermediate coupling regime.
These benchmarks demonstrate that the optimized polaron frame provides a compact and quantitatively reliable starting point for strongly coupled light--matter and electron--phonon ground states.

The resulting framework combines an asymptotically decoupled dressed representation, variational optimization, an explicit scan of the residual bosonic Hilbert space, and perturbative correction for residual entanglement. 
It therefore avoids both bare-state truncation and the transformed-vacuum restriction, providing a compact route for molecular polariton and polaron ground states.

\begin{figure*}
	\centering
	\includegraphics[width=0.9\linewidth]{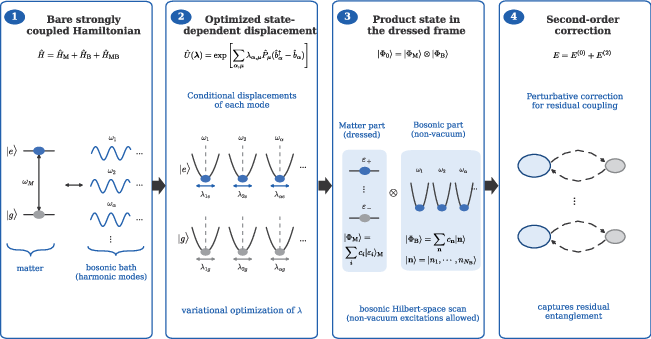}
	\caption{Schematic of the variational polaron theory. Optimized state-dependent displacements absorb the dominant bosonic dressing into the basis, yielding a dressed frame in which a matter--boson product state is variationally optimized. Residual matter--boson correlations are subsequently included by a second-order perturbative correction.}
	\label{fig: schematic workflow}
\end{figure*}

\textit{Variational polaron framework.}
We consider a matter-boson Hamiltonian 
$\hat{H}=\hat{H}_{\rm M}+\hat{H}_{\rm B}+\hat{H}_{\rm MB}$, where $\hat{H}_{\rm M}$ describes electronic, spin, or other matter degrees of freedom, and $\hat{H}_{\rm B}=\sum_\alpha\omega_\alpha\hat{b}_\alpha^\dagger\hat{b}_\alpha$ describes bosonic modes. 
For clarity, let us consider the common situation in which the dominant linear matter-boson coupling can be resolved in a set of orthogonal and complete matter projectors $\{\hat{P}_\mu\}$, 
\begin{align}
	\hat{P}_\mu\hat{P}_\nu=\delta_{\mu\nu}\hat{P}_\mu,
	\qquad
	\sum_\mu\hat{P}_\mu=\hat I_{\rm M},
\end{align}
with
\begin{align}
	\hat{H}_{\rm MB}=\sum_{\alpha,\mu} g_{\alpha,\mu}\hat{P}_\mu(\hat{b}_\alpha^\dagger+\hat{b}_\alpha),
\end{align}
possibly in addition to off-diagonal matter terms contained in $\hat{H}_{\rm M}$. 
We introduce the state-dependent variational polaron transformation
\begin{align}
	\hat{U}(\bm\lambda)=\exp\left[\sum_{\alpha,\mu}\lambda_{\alpha,\mu}\hat{P}_\mu
	(\hat{b}_{\alpha}^{\dagger}-\hat{b}_{\alpha})\right],
	\label{eq: general polaron transformation}
\end{align} 
where the displacement amplitudes $\lambda_{\alpha,\mu}$ are real variational parameters. 
The transformed Hamiltonian is $\hat{H}(\bm\lambda)=\hat{U}^\dagger(\bm\lambda)\hat{H}\hat{U}(\bm\lambda)$. 
Since 
\begin{align}
	\hat{U}^\dagger\hat{b}_\alpha\hat{U}
	=
	\hat{b}_\alpha+\sum_\mu \lambda_{\alpha,\mu}\hat{P}_\mu,
\end{align}
the transformed bosonic and matter-boson terms can be written as
\begin{align}
	\hat{U}^\dagger
	(\hat{H}_{\rm B}+\hat H_{\rm MB})
	\hat{U}
	=
	\hat H_{\rm B}
	+
	\hat H_{\rm res}
	+
	\sum_\mu E_\mu^{\rm shift}\hat P_\mu,
	\label{eq: transformed diagonal coupling}
\end{align}
where
\begin{align}
	\hat H_{\rm res}
	&=
	\sum_{\alpha,\mu}
	R_{\alpha,\mu}\hat{P}_\mu
	(\hat{b}_\alpha^\dagger+\hat{b}_\alpha),
	\\
	R_{\alpha,\mu}
	&=
	g_{\alpha,\mu}+\omega_\alpha\lambda_{\alpha,\mu},
	\\
	E_\mu^{\rm shift}
	&=
	\sum_\alpha
	\left(
	\omega_\alpha\lambda_{\alpha,\mu}^2
	+2g_{\alpha,\mu}\lambda_{\alpha,\mu}
	\right).
\end{align}
Thus, the optimal strong-coupling displacement approaches the value $\lambda_{\alpha,\mu}\simeq -g_{\alpha,\mu}/\omega_\alpha$ that cancels the residual linear coupling $\hat H_{\rm res}$, up to the sign convention of Eq.~\eqref{eq: general polaron transformation}.

The decoupling of off-diagonal matter terms follows from the block structure of Eq.~\eqref{eq: general polaron transformation}.
Defining
\begin{align}
	\hat{A}_\mu
	=
	\sum_\alpha
	\lambda_{\alpha,\mu}
	(\hat b_\alpha^\dagger-\hat b_\alpha),
\end{align}
the transformation can be written as
\begin{align}
	\hat{U}
	=
	\sum_\mu \hat{P}_\mu e^{\hat A_\mu},
	\qquad
	\hat{U}^\dagger
	=
	\sum_\mu \hat{P}_\mu e^{-\hat A_\mu}.
\end{align}
Matter operators that are diagonal with respect to $\{\hat P_\mu\}$ are unchanged by the transformation: for $\hat O_{\rm diag}=\sum_\mu\hat P_\mu\hat O\hat P_\mu$, one has $\hat U^\dagger\hat O_{\rm diag}\hat U=\hat O_{\rm diag}$.
In contrast, an off-diagonal matter operator $\hat{O}_{\mu\nu}=\hat{P}_\mu \hat{O} \hat{P}_\nu$ transforms as \begin{align}
	\hat{U}^\dagger \hat{O}_{\mu\nu} \hat{U}
	=
	\hat{O}_{\mu\nu}
	\prod_\alpha
	\exp\left[ 
	(\lambda_{\alpha,\nu}-\lambda_{\alpha,\mu})
	(\hat{b}_\alpha^\dagger-\hat{b}_\alpha)
	\right].
\end{align}
with the detailed derivation given in the Supporting Information. 
Thus, off-diagonal matter transitions are multiplied by displaced-oscillator overlaps. 
Their matrix elements between Fock states contain
\begin{align}
	D_{nm}^\lambda
	=
	\langle n|e^{\lambda(\hat b^\dagger-\hat b)}|m\rangle.
\end{align}
For example, for $n\ge m$, 
\begin{align}
	D_{nm}^\lambda
	=
	\sqrt{\frac{m!}{n!}}\,
	e^{-\lambda^2/2}
	\lambda^{n-m}
	L_m^{n-m}(\lambda^2),
\end{align}
with an analogous expression for $n\le m$. 
Thus, for fixed finite $n$ and $m$, $D_{nm}^\lambda\to 0$ as $|\lambda|\to\infty$, because the associated Laguerre polynomial grows only algebraically whereas the factor $e^{-\lambda^2/2}$ decays exponentially.
Consequently, when the displacement differences $\lambda_{\alpha,\nu}-\lambda_{\alpha,\mu}$ grow with the coupling strength, the off-diagonal residual matter-boson matrix elements vanish within any fixed finite bosonic sector. 
At the strong-coupling optimum, the transformed Hamiltonian therefore approaches 
\begin{align}
	\hat H(\bm\lambda)
	\longrightarrow
	\hat H_{\rm B}
	+
	\sum_\mu
	\hat P_\mu\hat H_{\rm M}\hat P_\mu
	+
	\sum_\mu
	E_\mu^{\rm shift}\hat P_\mu,
	\nonumber\\
	\qquad
	|g|/\omega\rightarrow\infty,
	\label{eq: asymptotically decoupled Hamiltonian}
\end{align}
where $\hat H_{\rm B}=\sum_\alpha\omega_\alpha\hat b_\alpha^\dagger\hat b_\alpha$ and $E_\mu^{\rm shift}\to -\sum_\alpha g_{\alpha,\mu}^2/\omega_\alpha$. 
Equation~\eqref{eq: asymptotically decoupled Hamiltonian} contains only a boson-only Hamiltonian and matter-only block-diagonal terms, with no residual product of matter and bosonic operators. 
This is the asymptotic decoupling of the transformed Hamiltonian in the infinite-coupling limit.

The zeroth-order ground state is taken as a product state between the matter and bosonic sectors in the transformed frame,
\begin{align}
	\ket{\Phi_0}=\ket{\Phi_{\rm M}}\otimes\ket{\Phi_{\rm B}}.
	\label{eq:product state}
\end{align}
This factorized form is motivated by the asymptotic decoupling of the optimized transformed Hamiltonian in the strong-coupling limit. 
After the dominant matter-induced bosonic deformation is absorbed into the state-dependent displacement, the residual matter-boson coupling is expected to become perturbative in the optimized dressed frame. 
Thus, at finite coupling, Eq.~\eqref{eq:product state} should be regarded as a zeroth-order dressed reference rather than as a final mean-field approximation. 

The matter coefficients, the bosonic coefficients, and the displacement parameters are optimized simultaneously by minimizing
\begin{align}
	E_{\rm var}=\bra{\Phi_0}\hat{H}(\bm\lambda)\ket{\Phi_0},
	\label{eq:evar}
\end{align}
where the trial state is normalized. 
Importantly, the factorization in Eq.~\eqref{eq:product state} is imposed only between the matter and bosonic sectors. 
The bosonic component is not fixed to the transformed oscillator vacuum and is not assumed to factorize over independent modes. 
Instead, it is expanded as a general multimode bosonic wave function in the retained Fock space,
\begin{align}
	\ket{\Phi_{\rm B}}
	=
	\sum_{\bm n}
	c_{\bm n}
	\ket{\bm n},
	\qquad
	\ket{\bm n}
	=
	\ket{n_1,\ldots,n_{N_{\rm B}}}.
	\label{eq: bosonic multimode state}
\end{align}
This allows the zeroth-order state to capture non-vacuum bosonic excitations and correlations between bosonic modes that remain in the optimized transformed frame, especially at intermediate coupling. 
The product-state structure is also useful computationally because the matter sector can, in principle, be treated independently using methods appropriate for the underlying matter problem, while the bosonic dressing is optimized in the transformed frame. 

Since the zeroth-order ansatz is still factorized between matter and bosons, it does not contain residual matter-boson entanglement beyond the optimized dressed-frame reference. 
To incorporate this remaining entanglement, we construct a mean-field zeroth-order Hamiltonian $\hat{H}^{(0)}(\bm \lambda)$ for which $\ket{\Phi_0}$ is the ground state and define 
\begin{align}
	\delta \hat{H}
	=
	\hat{H}(\bm \lambda)
	-
	\hat{H}^{(0)}(\bm \lambda).
\end{align}
The second-order energy correction is then given by
\begin{align}
	E^{(2)}
	=
	\sum_{n\ne0}
	\frac{|\bra{\Phi_n}\delta\hat{H}\ket{\Phi_0}|^2}{E_0-E_n},
	\label{eq: second-order energy correction}
\end{align}
where $\ket{\Phi_n}$ are excited eigenstates of $\hat{H}^{(0)}$ with energies $E_n$. 
The corresponding first-order correction to the wave function is used to evaluate the corrected fidelity. 

\textit{Collective light-matter benchmark: Dicke model.}
We first apply the method to the Dicke Hamiltonian,
\begin{align}
	\hat{H}_{\rm D}=\omega\hat{a}^\dagger\hat{a}+\omega_0\hat{S}_z+g(\hat{a}^\dagger+\hat{a})\hat{S}_x,
	\label{eq: Dicke Hamiltonian}
\end{align}
where $\hat{S}_\alpha=\frac{1}{2}\sum_{j=1}^N\hat{\sigma}_\alpha^{(j)}$ ($\alpha=x,y,z$) are collective spin operators for $N$ two-level emitters. 
The Dicke model is used here as a minimal benchmark for collective light-matter coupling. 
It is not a complete molecular polariton Hamiltonian, because it does not explicitly include nuclear or vibronic structure, but it provides a stringent test of counter-rotating coupling, collective dressing, and the superradiant quantum phase transition~\cite{Dicke1954,HeppLieb1973,WangHioe1973}.

We choose projectors $\hat{P}_l=\ket{u_l}\bra{u_l}$ onto eigenstates of $\hat{S}_x$, with eigenvalue $l=-N/2,\cdots,N/2$, and use
\begin{align}
	\hat{U}_{\rm D}(\bm\lambda)=\exp\left[(\hat{a}^\dagger-\hat{a})\sum_l\lambda_l\hat{P}_l\right].
	\label{eq: Dicke polaron transformation U}
\end{align}
This is the direct specialization of the general projector-dependent displacement to the collective spin coordinate. 
In sector $l$, the cavity mode experiences a force proportional to $gl$; the optimized displacement therefore removes this force in the strong-coupling limit, whereas the remaining spin terms that connect different $l$ sectors acquire displaced-oscillator overlaps. 
The Dicke benchmark consequently tests both ingredients of the theory: cancellation of the dominant diagonal light-matter coupling and suppression of residual off-diagonal transitions in the dressed frame.
The zeroth-order state is written as 
\begin{align}
	\ket{\Phi_0}=\left(\sum_l A_l\ket{u_l}\right)\otimes
	\left(\sum_{n=0}^{n_{\rm max}}c_n\ket{n}\right),
	\label{eq: Dicke product state ansatz}
\end{align}
and the coefficients $\{A_l,c_n\}$ and displacements $\{\lambda_l\}$ are optimized by gradient descent.
The expansion over photon Fock states in Eq.~\eqref{eq: Dicke product state ansatz} is important because the transformed-frame photon state is not assumed to be the vacuum; residual non-vacuum components are optimized explicitly.
Matrix elements of $\hat{H}_{\rm D}(\bm \lambda)$ contain overlaps between displaced number states and are evaluated analytically using the Laguerre-polynomial formula discussed above and in the Supporting Information. 
In the large-$g$ limit, these overlaps vanish for fixed finite Fock states whenever two spin sectors have different displacements, so the transformed Hamiltonian becomes asymptotically decoupled.

\begin{figure}
	\centering
	\includegraphics[width=0.9\linewidth]{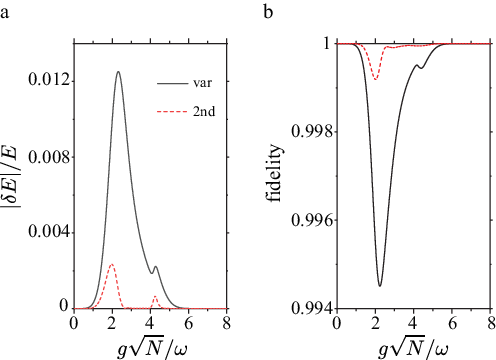}
	\caption{
			Dicke-model ground-state accuracy for $N=1$. 
			`var' denotes the optimized zeroth-order result, whereas `2nd' includes the second-order correction.
			The correction lowers the maximum energy error below $0.2\%$ and raises the fidelity above $0.999$, quantifying the practical onset of transformed-frame decoupling.
		}

	\label{fig: Dicke accuracy}
\end{figure}

Figure~\ref{fig: Dicke accuracy} summarizes the Dicke benchmark for $N=1$. 
The zeroth-order variational calculation already gives a maximum relative energy error of about $1.2\%$, while the second-order correction reduces the maximum error to about $0.2\%$. 
The fidelity is similarly improved: the minimum fidelity is approximately $0.995$ at the variational level and exceeds $0.999$ after the perturbative correction. 
The errors vanish in both $g\to 0$ and $g\to \infty$ limits, as expected from the construction, and peak only in the intermediate regime where neither a bare-state nor a fixed strong-coupling description is optimal. 
Thus, the benchmark does more than confirm the formal asymptotic limit: it quantifies how rapidly the optimized transformed frame approaches effective decoupling as a function of the dimensionless collective coupling and gives the residual energy and wave-function errors that remain at intermediate coupling. 

\begin{figure}
	\centering
	\includegraphics[width=0.9\linewidth]{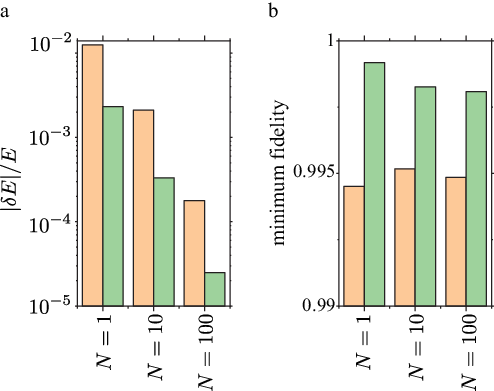}
	\caption{
			Scaling of Dicke-model accuracy with emitter number.
			(a) Maximum relative ground-state energy error, $|\delta E|/E$, for $N=1$, $10$, and $100$; orange and green bars denote the zeroth-order and second-order results, respectively.
			(b) Minimum ground-state fidelity over the same coupling range, showing near-unity agreement after the perturbative correction.
		}

	\label{fig: Dicke compare energy error and fidelity}
\end{figure}

The method also improves with system size, as shown in Fig.~\ref{fig: Dicke compare energy error and fidelity}. 
As the number of emitters increases, the relative energy error decreases from the order of $10^{-2}$ for $N=1$ to $10^{-3}$ for $N=10$ and $10^{-4}$ for $N=100$, while the fidelity remains larger than $0.998$ across the tested values of $N$. 
This improvement is consistent with the collective structure of the Dicke coupling. 
The optimized displacement absorbs the dominant photon shift associated with each $S_x$ sector, so that the remaining transformed-frame coupling arises mainly from transitions between differently displaced sectors. 
These residual transitions are suppressed by displaced-oscillator overlaps and therefore become increasingly amenable to perturbative treatment in the collective regime.
The benchmark therefore identifies not only the formal infinite-coupling limit, but also the finite-$N$ and finite-coupling accuracy of the dressed-frame approximation.

\begin{figure}
	\centering
	\includegraphics[width=0.9\linewidth]{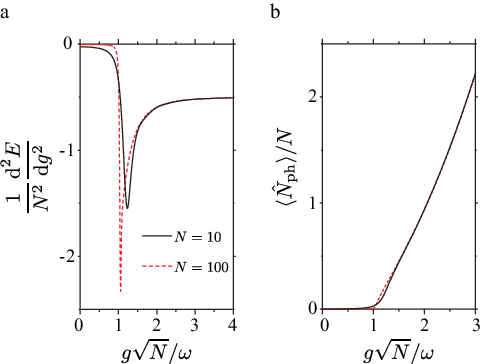}
	\caption{
			Superradiant quantum phase transition in the Dicke model.
			(a) Scaled second derivative of the ground-state energy, $N^{-2}d^2E/dg^2$, versus $g\sqrt{N}/\omega$ for $N=10$ and $N=100$.
			The dip sharpens toward the critical coupling $g\sqrt{N}/\omega=1$.
			(b) Photon occupation per emitter, $\langle \hat{N}_{\rm ph}\rangle/N$, confirming the onset of the superradiant phase.
		}

	\label{fig: Dicke energy's 2nd derivative and photon condensation}
\end{figure}

A further test is the superradiant phase transition. 
In the thermodynamic limit, the Dicke model undergoes a second-order transition from the normal phase to a superradiant phase with macroscopic photon occupation. 
As shown in Fig.~\ref{fig: Dicke energy's 2nd derivative and photon condensation}, the optimized polaron theory reproduces the sharp feature in the second derivative of the ground-state energy and the onset of photon condensation at the expected critical coupling. 
We emphasize that the Dicke Hamiltonian is used here as a benchmark model; applications to gauge-invariant molecular cavity Hamiltonians should include the corresponding dipole self-energy or diamagnetic terms, which can be incorporated within the same transformed-frame strategy~\cite{Rzazewski1975,Stokes2019}. 

\textit{Electron-phonon benchmark: Holstein model.}
We next consider the one-electron Holstein Hamiltonian,
\begin{align}
	\hat{H}_{\rm H}=
	-J\sum_{\langle i,j\rangle}(\hat{c}_i^\dagger\hat{c}_j+\hat{c}_j^\dagger\hat{c}_i)
	+\omega\sum_i\hat{b}_i^\dagger\hat{b}_i
	+g\sum_i\hat{n}_i(\hat{b}_i^\dagger+\hat{b}_i),
	\label{eq: Holstein model Hamiltonian}
\end{align}
where $\hat{n}_i=\hat{c}_i^\dagger\hat{c}_i$.
This model is directly relevant to the competition between electronic delocalization and local vibrational dressing in molecular crystals, organic semiconductors, and related polaronic materials~\cite{Holstein1959,Alexandrov2010}. 
We use the variational transformation
\begin{align}
	\hat{U}_{\rm H}(\bm\lambda)=\exp\left[\sum_i\lambda_i\hat{n}_i(\hat{b}_i^\dagger-\hat{b}_i)\right],
	\label{eq: Holstein polaron transformation}
\end{align}
where $\lambda_i$ describes the phonon displacement at site $i$ generated by an electron at the same site. 
In this site-local projector basis, the electron-phonon term is diagonal, while the hopping operator is off-diagonal because it transfers the electron between different displacement sectors. 
After the transformation, hopping is therefore multiplied by phonon displacement overlaps between the two sites involved in the hop. 
In the strong-coupling limit these overlaps suppress hopping between inequivalent local distortions, giving the usual localized-polaron picture, whereas at intermediate coupling the optimized Fock-space bosonic state captures residual phonon excitations beyond a pure transformed vacuum.
Translational symmetry can be imposed by requiring $\lambda_i$ to be site-independent, or it can be relaxed to test whether symmetry breaking improves the variational state. 

\begin{figure}
	\centering
	\includegraphics[width=0.9\linewidth]{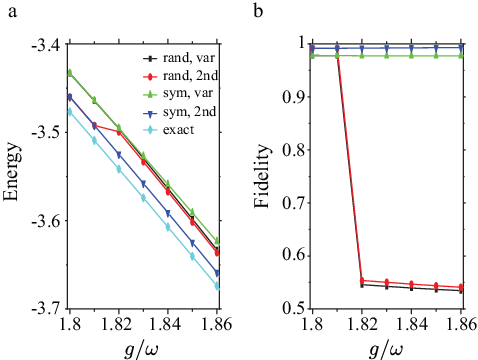}
	\caption{
			Effect of translational-symmetry constraint on the Holstein-model ground state near intermediate coupling.
			(a) Ground-state energy versus $g/\omega$ for unconstrained random initial states (`rand') and translationally symmetric states (`sym'), compared with exact diagonalization; `var' and `2nd' denote zeroth-order and second-order results.
			(b) Corresponding fidelity. 
			The unconstrained solution can have lower variational energy but poor fidelity, whereas the symmetry-constrained branch remains close to the exact ground state.
		}

	\label{fig: Holstein random vs fixed initial state energy and fidelity}
\end{figure}

The Holstein benchmark is shown in Fig.~\ref{fig: Holstein random vs fixed initial state energy and fidelity}.
The numerical calculations were performed for a two-site single-electron Holstein model with $J/\omega=1$, while the electron--phonon coupling strength $g/\omega$ was varied. 
This minimal setting captures the competition between electronic delocalization and local phonon dressing and provides a stringent test of the optimized transformed-frame description in the polaron-crossover regime.
The maximum energy error is approximately $1.4\%$ at the variational level and decreases to approximately $0.5\%$ after the second-order correction. 
As in the Dicke model, the errors vanish in both the weak- and strong-coupling limits, showing that the same variational principle captures both delocalized electronic motion and localized phonon dressing. 
The remaining error and fidelity provide a direct measure of the coupling range over which residual matter-boson correlations remain important, thereby identifying the practical onset of the effectively decoupled polaron frame for a chosen numerical tolerance. 

The calculation also reveals a subtle role of translational symmetry. 
Relaxing the symmetry constraint can lower the zeroth-order variational energy at intermediate-to-strong coupling, but the fidelity then drops sharply. 
This discontinuity signals a switch between competing variational branches: in the present translationally invariant benchmark, the unconstrained branch is overlocalized and symmetry-broken, whereas the symmetry-constrained branch remains connected to the exact ground state and gives the better corrected energy. 
Thus, a lower uncorrected variational energy does not necessarily imply a better physical state when residual matter-boson correlations and symmetry restoration are important. 
In systems without translational symmetry, however, such branch changes may represent physical localization. 
Continuity under parameter changes, stability against initial conditions, small perturbative corrections, and agreement with observables then provide more robust diagnostics than the zeroth-order energy alone.

\textit{Implications for polariton and polaron physics and chemistry.}
The results show that the optimized state-dependent polaron transformation provides a compact dressed basis across weak-, intermediate-, and strong-coupling regimes. 
In the weak-coupling limit the displacement vanishes, while in the strong-coupling limit the transformed Hamiltonian becomes asymptotically decoupled because the dominant diagonal matter-induced bosonic deformation is absorbed and off-diagonal matter-sector transitions are suppressed by displaced-oscillator overlaps. 
This behavior explains why the errors vanish in both limits and why the intermediate regime is the most stringent test.

The asymptotic decoupling property also clarifies the role of the product-state ansatz. 
The zeroth-order state is not a bare mean-field approximation, but a dressed reference constructed after the dominant matter-boson dressing has been absorbed into the basis. 
At finite coupling, residual transformed-frame couplings remain, but the benchmarks show that their leading energetic effect can be captured efficiently by the second-order correction. 
The reliability of the product reference can therefore be diagnosed by the residual coupling, the perturbative correction, and the fidelity.

The formulation is also modular. 
Because the zeroth-order state factorizes only between the matter and bosonic sectors, the matter component can in principle be treated using methods tailored to the underlying matter problem, while the bosonic dressing is optimized through the state-dependent displacement and retained multimode Fock space. 
The bosonic wave function is not restricted to the transformed vacuum or to independent modes, so non-vacuum multimode correlations can be retained. 
This separation suggests a route for combining the present transformation with configuration-interaction, tensor-network, electronic-structure, or model-Hamiltonian solvers for the matter sector, while correcting residual matter-boson entanglement perturbatively.

These features are relevant to both polariton and polaron problems. 
The optimized dressed basis may provide a starting point for multimode molecular polariton ground states, structured electron-phonon environments, and future dynamical or spectroscopic calculations, including coherent multidimensional probes of many-body interactions~\cite{Phuc2021PRB}, polariton-assisted energy or charge transfer, and spin-selective transport.
More broadly, the framework bridges nonperturbative basis optimization and perturbative corrections for strongly dressed ground states in cavity-modified chemistry and molecular materials.

\textit{Conclusion} 
We have introduced a variational polaron theory for ground states of strongly coupled matter-boson systems. 
The method combines a state-dependent polaron transformation, an optimized matter-boson product state, a general multimode bosonic wave function in the retained Fock space, and a second-order correction for residual matter-boson entanglement. 
The optimized transformed Hamiltonian becomes asymptotically decoupled in the strong-coupling limit, providing both a physical basis for the dressed product reference and a diagnostic for when a compact transformed-frame description is reliable.

Benchmarks on the Dicke and Holstein models show that the same framework captures collective light--matter dressing, the superradiant transition, and electron--phonon polaron formation with high accuracy. 
The second-order correction consistently improves the zeroth-order result by accounting for residual matter--boson correlations, while the Holstein benchmark highlights the importance of symmetry constraints for wave-function quality.

Because the approach separates the matter and bosonic sectors at the zeroth-order level while retaining a correlated multimode bosonic wave function, it is naturally compatible with problem-specific matter solvers. 
This dressed-basis strategy should be useful for multimode polariton ground states, structured electron--phonon environments, and first-principles models of strongly coupled molecular and materials systems.

\begin{acknowledgments}
N. T. P acknowledges financial support from the JST PRESTO program ``Exploring Quantum Frontiers Through Quantum-Classical Interdisciplinary Fusion'' (JPMJPR24F2) and the Hirose Foundation.
The computational work was conducted using the facilities of the Research Center for Computational Science, Okazaki, Japan.
\end{acknowledgments}

\section*{Supporting Information Available}
Derivations of the transformed Dicke and Holstein Hamiltonians, variational energy expressions, displaced-state overlap matrix elements, asymptotic decoupling limits, and perturbative corrections.

\section*{Notes}
The author declares no competing financial interest.






\end{document}